\documentclass[%
 aip,
 floatfix,
 amsmath,amssymb,
 reprint,%
]{revtex4-1}

\usepackage{nicematrix}
\usepackage{mathtools}
\usepackage{graphicx}
\usepackage{dcolumn}
\usepackage{bm}

\newcommand{\pderiv}[2]{
\frac{\partial #1}{\partial #2}
}
\newcommand{\pderivInline}[2]{
\partial #1/\partial #2
}

\usepackage{graphicx}
\usepackage{dcolumn}
\usepackage{bm}

\usepackage[utf8]{inputenc}
\usepackage[T1]{fontenc}
\usepackage{mathptmx}
\usepackage{etoolbox}

\usepackage[utf8]{inputenc} 
\usepackage{url}
\usepackage{amsmath}
\usepackage{amssymb}    
\usepackage{graphicx}
\usepackage{listings}
\usepackage{hyperref}

\renewcommand{\v}[1]{\ensuremath{\mathbf{#1}}} 
\newcommand{\gv}[1]{\ensuremath{\mbox{\boldmath$ #1 $}}} 
\newcommand{\grad}[1]{\nabla #1} 
\renewcommand{\div}[1]{\nabla \cdot #1} 
\newcommand{\pd}[2]{\frac{\partial #1}{\partial #2}} 
\newcommand{\uv}[1]{\ensuremath{\mathbf{\hat{#1}}}} 
 
\renewcommand{\d}[2]{\frac{d #1}{d #2}} 
\newcommand{\delpar}[1]{\nabla_{\parallel} #1} 
\newcommand{\gradperp}[1]{\grad{}_{\perp} {#1}} 

\newcommand{\curv}[1]{{C}_{\left({#1}\right)}}

\newcommand{\cs}{c_s}

\newcommand{\bhatZ}{\v{b_0}}

\newcommand{\cur}{j_{\parallel}}

\newcommand{\vpi}{v_{\parallel i}}						
\newcommand{\vpe}{v_{\parallel e}}					
\newcommand{\gvort}{\omega}		

\newcommand{\momSrce}{S_{\mathcal{M}\parallel e}}							
\newcommand{\momSrci}{S_{\mathcal{M}\parallel i}}							
\newcommand{\nSrcN}{S_n}									
\newcommand{\enerSrceN}{S_{E,e}}							
\newcommand{\enerSrciN}{S_{E,i}}							

\newenvironment{eqnal}{\equation\aligned}{\endaligned\endequation}

\newcommand{\fortranl}[1]{\lstinline[language={[90]Fortran}]{#1}}

\makeatletter
\def\@email#1#2{%
 \endgroup
 \patchcmd{\titleblock@produce}
  {\frontmatter@RRAPformat}
  {\frontmatter@RRAPformat{\produce@RRAP{*#1\href{mailto:#2}{#2}}}\frontmatter@RRAPformat}
  {}{}
}%
\makeatother
\begin{document}

\preprint{AIP/123-QED}

\title{Turbulent field fluctuations in gyrokinetic and fluid plasmas}
\author{A. Mathews}
 \email{mathewsa@mit.edu}
\affiliation{MIT Plasma Science and Fusion Center, Cambridge, Massachusetts 02139, USA
}
\author{N. Mandell} 
\affiliation{MIT Plasma Science and Fusion Center, Cambridge, Massachusetts 02139, USA
}
\author{M. Francisquez}
\affiliation{Princeton Plasma Physics Laboratory, Princeton, New Jersey 08540, USA}
\author{J.W. Hughes} 
\affiliation{MIT Plasma Science and Fusion Center, Cambridge, Massachusetts 02139, USA
}
\author{A. Hakim}
\affiliation{Princeton Plasma Physics Laboratory, Princeton, New Jersey 08540, USA}

\date{\today}

\begin{abstract}
A key uncertainty in the design and development of magnetic confinement fusion energy reactors is predicting edge plasma turbulence. An essential step in overcoming this uncertainty is the validation in accuracy of reduced turbulent transport models. Drift-reduced Braginskii two-fluid theory is one such set of reduced equations that has for decades simulated boundary plasmas in experiment, but significant questions exist regarding its predictive ability. To this end, using a novel physics-informed deep learning framework, we demonstrate the first ever direct quantitative comparisons of turbulent field fluctuations between electrostatic two-fluid theory and electromagnetic gyrokinetic modelling with good overall agreement found in magnetized helical plasmas at low normalized pressure. This framework presents a new technique for the numerical validation and discovery of reduced global plasma turbulence models.
\end{abstract}

\maketitle

\section{\label{sec:level1}Introduction}

A hallmark of a turbulence theory is the nonlinear connection that it sets between dynamical variables. For ionized gases, perturbations in the plasma are intrinsically linked with perturbations of the electromagnetic field. Accordingly, the electric potential response concomitant with electron pressure fluctuations constitutes one of the key relationships demarcating a plasma turbulence theory and presents a significant consistency test en route to its validation. But to meaningfully evaluate this dynamical interplay across theories or experiment using standard numerical methods is an intricate task. Namely, when comparing global plasma simulations to other numerical models or diagnostics, one must precisely align initial states, boundary conditions, and sources of particles, momentum, and energy (e.g. plasma heating, sheath effects, atomic and molecular processes) \cite{francisquez2020fluid,Mijin_2020}. These descriptions of physics can be increasingly difficult to reconcile when applying different representations---for example, fluid or gyrokinetic---or when measurements characterizing the plasma are sparse as commonly encountered in thermonuclear fusion devices. Additionally, imperfect conservation properties (both in theory and numerics due to discretization) and the limited accuracy of time-integration schema can introduce errors further misaligning comparative modelling efforts. Such difficulties exist even when examining the same nominal turbulence theory across numerical codes \cite{nonlinear_GK_comparison_new,Tronko_ORB5vGENE,nonlinear_GK_comparison1,nonlinear_GK_comparison2} and become magnified for cross-model comparisons \cite{francisquez2020fluid}. To overcome these classical limitations, we apply a custom physics-informed deep learning framework \cite{mathews2020uncovering} with a continuous spatiotemporal domain and partial electron pressure observations to demonstrate the first direct comparisons of instantaneous turbulent fields between two distinct global plasma turbulence models: electrostatic drift-reduced Braginskii theory and electromagnetic long-wavelength gyrokinetics.

This is significant since validating the accuracy of reduced turbulence theories is amongst the greatest challenges to the advancement of plasma modelling. In the fusion community, both comprehensive full-$f$ global gyrokinetic \cite{mandell_hakim_hammett_francisquez_2020,XGC,GENE,COGENT,ELMFIRE,ORB5,GYSELA,PICLS} and two-fluid \cite{BOUT++,TOKAM3X,GBS,GRILLIX,GDB} direct numerical simulations are under active development to improve predictive modelling capabilities and the design of future reactors, but no single code can currently capture all the dynamics present in the edge with sufficient numerical realism. It is therefore of paramount importance to recognize when certain numerical or analytical simplifications make no discernible difference, or when a given theoretical assumption is too risky---to clearly identify the limits of present simulations \cite{francisquez2020fluid}. Validating reduced plasma turbulence models is accordingly essential. Adaptations of the drift-reduced Braginskii equations have recently investigated several important edge phenomena \cite{Francisquez_2017,Ricci_2012,Thrysoe_2018,nespoli_non-linear_2017}, but precise quantitative agreement with experiment is generally lacking on a wide scale and their accuracy in fusion reactors is uncertain, especially since kinetic effects may not be negligible as edge plasma dynamics approach the ion Larmor radius and collision frequencies are comparable to turbulent timescales \cite{Braginskii1965RvPP,blobs_review}. Yet these fluid codes remain in widespread usage for their reduced computational cost, which offers the ability to perform parameter scans with direct numerical simulations that can help uncover underlying physics and optimize reactor engineering designs. Nevertheless, the Vlasov-Maxwell system of equations is the gold standard in plasma theory, but such modelling is prohibitively expensive when wholly simulating the extreme spatiotemporal scales extant in fusion reactors. For the purposes of more tractable yet high fidelity numerical simulations, this system is averaged over the fast gyro-motion of particles to formulate 5D gyrokinetics. Fluid-gyrokinetic comparisons therefore present a way to analyze the robustness of reduced modelling while also illuminating when collisional fluid theories may be insufficient. Quantifying potential shortcomings is thus vital for the testing and development of improved reduced models \cite{waltz2019kinetic}. 

A particularly important quantity for testing predictive modelling is edge electric field fluctuations since turbulent ${\bf E \times B}$ drifts strongly influence edge plasma stability and transport of blob-like structures which can account for over 50\% of particle losses during standard operation of tokamaks \cite{blobs_review,Simakov-paper,osti_1630104, osti_1762124,Carralero_2017,Kube_2018}. Resultant turbulent fluxes impacting the walls can cause sputtering, erosion, and impurity injection which adversely affect safe operation and confinement of fusion plasmas \cite{blobs_review,kuang_SPARC,kuang_ARC}. While explicit comparisons of global nonlinear dynamics across turbulence models were previously impractical using standard numerical simulations, our physics-informed deep learning \cite{raissi_PINN} technique enables direct quantitative cross-examination of plasma turbulence theories by uncovering the turbulent electric field expected in two-fluid and gyrokinetic models based upon an electron pressure fluctuation \cite{mathews2020uncovering}. This validation framework is potentially transferable to magnetized plasma turbulence in contexts beyond fusion energy such as space propulsion systems and astronomical settings as well. Overall, these physics-informed neural networks present a new paradigm in the advancement of reduced plasma turbulence models.

To demonstrate these results, we proceed with a description of gyrokinetic modelling based upon \citet{mandell_hakim_hammett_francisquez_2020} in Section \ref{sec:level2}, outline a custom physics-informed machine learning architecture from \citet{mathews2020uncovering} suited for the analysis of two-fluid theory in Section \ref{sec:level3}, present the first direct comparisons of instantaneous turbulent fields in section \ref{sec:level4}, and conclude with a summary and future outlook in Section \ref{sec:level5}.

\section{\label{sec:level2} Gyrokinetic modelling}

Following the full treatment outlined in \citet{mandell_thesis} and \citet{mandell_hakim_hammett_francisquez_2020}, the long-wavelength gyrokinetic modelling analyzed constitutes nonlinear global electromagnetic turbulence simulations on open field lines using the \texttt{Gkeyll} computational plasma framework. The full-$f$ electromagnetic gyrokinetic equation is solved in the symplectic formulation \cite{brizard2007foundations}, which describes the evolution of the gyrocenter distribution function $f_s(\v{Z},t)=f_s(\v{R},v_\parallel,\mu,t)$ for each species ($s = \{e,i\}$), where $\v{Z}$ is a phase-space coordinate composed of the guiding center position $\v{R}=(x,y,z)$, parallel velocity $v_\parallel$, and magnetic moment $\mu=m_s v_\perp^2/2B$. In terms of the gyrocentre Hamiltonian and the Poisson bracket in gyrocentre coordinates with collisions $C[f_s]$ and sources $S_s$, the gyrokinetic equation is
\begin{align}
    \pderiv{f_s}{t} + \{f_s, H_s\} - \frac{q_s}{m_s}\pderiv{A_\parallel}{t}\pderiv{f_s}{v_\parallel} = C[f_s] + S_s, \label{liouville}
\end{align}
or, equivalently,
\begin{align}
    \pderiv{f_s}{t} + \dot{\v{R}}\v{\cdot}\nabla f_s + \dot{v}^H_\parallel \pderiv{f_s}{v_\parallel}- \frac{q_s}{m_s}\pderiv{A_\parallel}{t}\pderiv{f_s}{v_\parallel} = C[f_s] + S_s,
\end{align}
where the gyrokinetic Poisson bracket is defined as
\begin{equation}
  \{F,G\} = \frac{\v{B^*}}{m_s B_\parallel^*} \v{\cdot} \left(\nabla F \frac{\partial G}{\partial v_\parallel} - \frac{\partial F}{\partial v_\parallel}\nabla G\right) - \frac{\uv{b}}{q_s B_\parallel^*}\times \nabla F \v{\cdot} \nabla G, 
\end{equation}
and the gyrocentre Hamiltonian is
\begin{equation}
    H_s = \frac{1}{2}m_sv_\parallel^2 + \mu B + q_s  \phi.
\end{equation}
The nonlinear phase-space characteristics are given by
\begin{equation}
    \dot{\v{R}} = \{\v{R},H_s\} = \frac{\v{B^*}}{B_\parallel^*}v_\parallel + \frac{\uv{b}}{q_s B_\parallel^*}\times\left(\mu\nabla B + q_s \nabla \phi\right),
\end{equation}
\begin{eqnal}\label{vpardot}
    \dot{v}_\parallel &= \dot{v}^H_\parallel -\frac{q_s}{m_s}\pderiv{A_\parallel}{t} = \{v_\parallel,H_s\}-\frac{q_s}{m_s}\pderiv{A_\parallel}{t} \\ \quad &= -\frac{\v{B^*}}{m_s B_\parallel^*}\v{\cdot}\left(\mu\nabla B + q_s \nabla \phi\right)-\frac{q_s}{m_s}\pderiv{A_\parallel}{t}.
\end{eqnal}

\noindent Here, $B_\parallel^*=\uv{b}\v{\cdot} \v{B^*}$ is the parallel component of the effective magnetic field $\v{B^*}=\v{B}+(m_s v_\parallel/q_s)\nabla\times\uv{b} + \delta \v{B}$, where $\v{B} = B \uv{b}$ is the equilibrium magnetic field and $\delta \v{B} = \nabla\times(A_\parallel \uv{b})\approx \nabla A_\parallel \times \uv{b}$ is the perturbed magnetic field {(assuming that the equilibrium magnetic field varies on spatial scales larger than perturbations so that $A_\parallel\nabla\times\uv{b}$ can be neglected)}. Higher-order parallel compressional fluctuations of the magnetic field are neglected so that $\delta\v{B}=\delta\v{B}_\perp$ and the electric field is given by ${\mathbf E} = -\nabla\phi - (\partial{A_\parallel}/\partial t) \uv{b}$. The species charge and mass are $q_s$ and $m_s$, respectively. In \eqref{vpardot}, $\dot{v}_\parallel$ has been separated into a term coming from the Hamiltonian, $\dot{v}^H_\parallel = \{v_\parallel,H_s\}$, and another term proportional to the inductive component of the parallel electric field, $(q_s/m_s)\partial {A_\parallel}/\partial{t}$. In the absence of collisions and sources, \eqref{liouville} can be identified as a Liouville equation demonstrating that the distribution function is conserved along the nonlinear phase space characteristics. Accordingly, the gyrokinetic equation can be recast in the following conservative form,
\begin{eqnal}
\pderiv{(\mathcal{J}f_s)}{t} &+ \nabla\v{\cdot}( \mathcal{J} \dot{\v{R}} f_s) + \pderiv{}{v_\parallel}\left(\mathcal{J}\dot{v}^H_\parallel f_s\right) \\ \quad & - \pderiv{}{v_\parallel}\left(\mathcal{J}\frac{q_s}{m_s}\pderiv{A_\parallel}{t} f_s \right) = \mathcal{J} C[f_s] + \mathcal{J} S_s, \label{emgk} 
\end{eqnal}
where $\mathcal{J} = B_\parallel^*$ is the Jacobian of the gyrocentre coordinates and $\uv{b}\v{\cdot}\nabla\times\uv{b}\approx0$ so that $B_\parallel^*\approx B$. The symplectic formulation of electromagnetic gyrokinetics is utilized with the parallel velocity as an independent variable instead of the parallel canonical momentum $p_\parallel$ in the Hamiltonian formulation \cite{brizard2007foundations,hahm1988nonlinear}. We use this notation for convenience, and to explicitly display the time derivative of $A_\parallel$, which is characteristic of the symplectic formulation of electromagnetic gyrokinetics. The electrostatic potential, $\phi$, is determined by quasi-neutrality,
\begin{equation}
    \sigma_g + \sigma_\text{pol} = \sigma_g - \nabla\v{\cdot}\v{P} = 0,
\end{equation}
with the guiding centre charge density (neglecting gyroaveraging in the long-wavelength limit) given by
\begin{equation}
    \sigma_g = \sum_s q_s \int d\v{w}\ \mathcal{J} f_s.
\end{equation}
Here, $d\v{w}= 2\pi\, m_s^{-1} dv_\parallel\, d\mu = m_s^{-1} dv_\parallel\, d\mu \int d\alpha$ is the gyrocentre velocity-space volume element
$(d{\bf v}=m_s^{-1}dv_\parallel\, d\mu\, d\alpha\, \mathcal{J})$ with the gyroangle $\alpha$ integrated away and the Jacobian factored out (formally, $\mathcal{J}$ should also be included in $d\v{w}$). The polarization vector is then
\begin{eqnal}
    \v{P} &= -\sum_s \int d\v{w}\ \frac{m_s}{B^2}\mathcal{J}f_s \nabla_\perp \phi \\ \quad & \approx -\sum_s \frac{m_s n_{0s}}{B^2} \nabla_\perp \phi \equiv - \epsilon_\perp \nabla_\perp \phi,
\end{eqnal}
where $\nabla_\perp=\nabla-\uv{b}(\uv{b}\v{\cdot}\nabla)$ is the gradient orthogonal to the background magnetic field. A linearized time-independent polarization density, $n_0$, is assumed which is consistent with neglecting a second-order ${\bf E \times B}$ energy term in the Hamiltonian. Such an approximation in the scrape-off layer (SOL) is questionable due to the presence of large density fluctuations, although a linearized polarization density is commonly used in full-$f$ gyrokinetic simulations for computational efficiency and reflective of common numerical modelling practices \cite{ku2018fast,shi2019full,mandell_hakim_hammett_francisquez_2020}. Adding the nonlinear polarization density along with the second-order ${\bf E \times B}$ energy term in the Hamiltonian are improvements kept for future work. Consequently, the quasi-neutrality condition can be rewritten as the long-wavelength gyrokinetic Poisson equation,
\begin{equation}
    -\nabla \v{\cdot} \sum_s \frac{m_s n_{0s}}{B^2} \nabla_\perp \phi = \sum_s q_s \int d\v{w}\ \mathcal{J}f_s, \label{poisson}
\end{equation}
where, even in the long-wavelength limit with no gyroaveraging, the first-order polarization charge density on the left-hand side of \eqref{poisson} incorporates some finite Larmor radius (FLR) or $k_\perp$ effects in its calculation. It is worth emphasizing that this “long-wavelength” limit is a valid limit of the full-$f$ gyrokinetic derivation since care was taken to include the guiding-center components of the field perturbations at $O(1)$. Further, although one may think of this as a drift-kinetic limit, the presence of the linearized ion polarization term in the quasineutrality equation distinguishes the long-wavelength gyrokinetic model from versions of drift-kinetics that, for example, include the polarization drift in the equations of motion.

The parallel magnetic vector potential, $A_\parallel$, is determined by the parallel Amp\`ere equation,
\begin{equation}
    -\nabla_\perp^2 A_\parallel = \mu_0 \sum_s q_s m_s \int  v_\parallel \mathcal{J} f_s\,d\v{w}. \label{ampere1}
\end{equation}
Note that we can also take the time derivative of this equation to get a generalized Ohm's law which can be solved directly for $\pderivInline{A_\parallel}{t}$, the inductive component of the parallel electric field $E_\parallel$ \cite{reynders1993gyrokinetic, cummings1994gyrokinetic, chen2001gyrokinetic}:
\begin{equation}
    -\nabla_\perp^2 \pderiv{A_\parallel}{t} = \mu_0 \sum_s q_s m_s \int v_\parallel \pderiv{(\mathcal{J} f_s)}{t}\, d\v{w}.
\end{equation}
Writing the gyrokinetic equation as
\begin{equation}
    \pderiv{(\mathcal{J}f_s)}{t} = 
    \pderiv{(\mathcal{J}f_s)}{t}^\star + \pderiv{}{v_\parallel}\left(\mathcal{J} \frac{q_s}{m_s}\pderiv{A_\parallel}{t} f_s\right), \label{fstar}
\end{equation}
where $\partial{(\mathcal{J}f_s)^\star}/\partial{t}$ denotes all terms in the gyrokinetic equation (including sources and collisions) except $\pderivInline{A_\parallel}{t}$ term, Ohm's law can be rewritten after an integration by parts as 
\begin{align}
    &\left(-\nabla_\perp^2 + \sum_s \mu_0 q_s^2 \int\mathcal{J} f_s\, d\v{w}\right) \pderiv{A_\parallel}{t} 
    \notag \\&\qquad\qquad
    = \mu_0 \sum_s q_s m_s \int v_\parallel \pderiv{(\mathcal{J}f_s)}{t}^\star\,d\v{w}. \label{ohmstar}
\end{align}

To model collisions, we use a conservative Lenard--Bernstein (or Dougherty) operator \cite{Lenard1958plasma,Dougherty1964model,francisquez2020conservative},

\begin{eqnal} \label{eq:GkLBOEq}
\mathcal{J}C[f_s] &= \nu\left\lbrace\pderiv{}{v_\parallel}\left[\left(v_\parallel - u_\parallel\right)\mathcal{J} f_s+v_{th,s}^2\pderiv{(\mathcal{J} f_s)}{v_\parallel}\right]\right.\\&\left.\quad +\pderiv{}{\mu}\left[2\mu \mathcal{J} f_s+2\mu\frac{m_s}{B}v_{th,s}^2\pderiv{(\mathcal{J} f_s)}{\mu}\right]\right\rbrace,
\end{eqnal}
where $n_s u_\parallel^2 = \int d\v{w} \mathcal{J}v_\parallel^2f_s$, $3 n_s v_{th,s}^2 = 2\int d\v{w} \mathcal{J}\mu Bf_s/m_s$, $n_s u_\parallel = \int d\v{w} \mathcal{J} v_\parallel f_s$, $n_s = \int d\v{w} \mathcal{J} f_s$, and $T_s = m_s v_{th,s}^2$. This collision operator contains the effects of drag and pitch-angle scattering, and it conserves number, momentum, and energy density. Consistent with our present long-wavelength treatment of the gyrokinetic system, FLR effects are ignored. In this work, both like-species and cross-species collisions among electrons and ions are included. The collision frequency $\nu$ is kept velocity-independent, i.e. $\nu\neq\nu(v)$. Further details about this collision operator, including its conservation properties and discretization, can be found in \cite{francisquez2020conservative}.

To clarify the approximations undertaken in deriving the gyrokinetic model formulated above and its consequent effects on turbulent fields, the key assumptions are reviewed: The orderings in gyrokinetic theory that effectively reduce the full phase space's dimensionality are ${\omega}/{\Omega_s} \ll 1$ and $k_\perp/k_\parallel \gg 1$. These orderings express that the charged particle gyrofrequency ($\Omega_s = q_s B/m_s$) in a magnetic field is far greater than the characteristic frequencies of interest ($\omega$) with perpendicular wavenumbers ($k_\perp$) of Fourier modes being far larger than parallel wavenumbers ($k_\parallel$). Such properties are generally expected and observed for drift-wave turbulence in magnetically confined fusion plasmas where $\omega \ll \Omega_s$ \cite{Zweben_2007}. An additional “weak-flow” ordering \cite{Dimits_1992,Parra_Catto_2008,Dimits_2012} is applied where $v_{\bf E \times B}/v_{th,s} \approx k_\perp \rho_s q_s \phi/T_s \ll 1$ which allows for large amplitude perturbations. This approximation is also generally valid for electrons and deuterium ions in edge tokamak plasmas \cite{Belli_2018} as it assumes ${\bf E \times B}$ flows are far smaller than the thermal velocity, $v_{th,s}$, as observed in experiment \cite{blobs_review,Zweben_2015}. By constraining gradients of $\phi$ instead of $\phi$ itself, this weak-flow ordering simultaneously allows perturbations of order unity ($q_s \phi/T_s \sim 1$) at long wavelengths ($k_\perp \rho_s \ll 1$) and small perturbations ($q_s \phi/T_s \ll 1$) at short wavelengths ($k_\perp \rho_s \sim 1$) along with perturbations at intermediate scales. Alternatively, this approximation can be intuitively viewed as the potential energy variation around a gyro-orbit being small compared to the kinetic energy, $q_s\phi(\v{R} + {\bm \rho_s}) - q_s\phi(\v{R}) \approx q_s {\bm \rho_s} \cdot \nabla_\perp \phi \ll T_s$ \cite{mandell_thesis}. Here ${\bm \rho_s}$ is the gyroradius vector which points from the center of the gyro-orbit $\v{R}$ to the particle location $\v{x}$. To ensure consistency in the gyrokinetic model at higher order (although the guiding-centre limit is eventually taken in the continuum simulations, i.e. $k_\perp \rho_s \ll 1$), a strong-gradient ordering is also employed which assumes the background magnetic field varies slowly relative to edge profiles \cite{Zweben_2007}. As noted above, we have taken the long-wavelength limit and neglected variations of $\phi$ on the scale of the gyroradius. This yields guiding-center equations of motion which do not contain gyroaveraging operations. While extensions to a more complete gyrokinetic model are in progress, these contemporary modelling limitations are worth noting for the scope of the present results.

In accordance with \citet{mandell_hakim_hammett_francisquez_2020}, the gyrokinetic turbulence is simulated on helical, open field lines as a rough model of the tokamak scrape-off layer at NSTX-like parameters. A non-orthogonal, field-aligned geometry \cite{beer1995field} is employed for numerical modelling whereby $x$ is the radial coordinate, $z$ is the coordinate parallel to the field lines, and $y$ is the binormal coordinate which labels field lines at constant $x$ and $z$. These coordinates map to physical cylindrical coordinates ($R,\varphi,Z)$ via $R=x$, $\varphi=(y/\sin\theta+z\cos\theta)/R_c$, $Z=z\sin\theta$. The field-line pitch $\sin\theta=B_v/B$ is taken to be constant, with $B_v$ the vertical component of the magnetic field (analogous to the poloidal field in tokamaks), and $B$ the total magnitude of the background magnetic field. The open field lines strike material walls at the top and bottom of the domain consistent with the simple magnetized torus configuration studied experimentally via devices such as the Helimak \cite{gentle2008} and TORPEX \cite{fasoli2006}. This system without magnetic shear contains unfavorable magnetic curvature producing the interchange instability that drives edge turbulence. There is no good curvature region to produce conventional ballooning-mode structure in the current setup. Further, $R_c=R_0+a$ is the radius of curvature at the centre of the simulation domain, with $R_0$ the major radius and $a$ the minor radius. The curvature operator is

\begin{equation}
    (\nabla\times\uv{b})\v{\cdot}\nabla f(x,y,z)\approx \left[(\nabla\times\uv{b})\v{\cdot} \nabla y\right]\pderiv{f}{y} =-\frac{1}{x}\pderiv{f}{y} \label{eq:GkCurv},
\end{equation}
where $\v{B}=B_\text{axis}(R_0/R)\uv{e}_z$ in the last step and $B_\text{axis}$ is the magnetic field strength at the magnetic axis.
This geometry represents a {flux-tube-like domain} on the outboard strictly bad curvature side that wraps helically around the torus and terminates on conducting plates at each end in $z$. {Note that although the simulation is on a flux-tube-like domain, it is not performed in the local limit commonly applied in $\delta f$ gyrokinetic codes; instead, the simulations are effectively global as they include radial variation of the magnetic field and kinetic profiles.} The simulation box is centred at $(x,y,z)=(R_c,0,0)$ with dimensions $L_x=56\rho_{i0}\approx 16.6$ cm, $L_y=100\rho_{i0}\approx 29.1$ cm, and $L_z=L_p/\sin\theta=8.0$ m, where $\rho_{i0} = \sqrt{m_i T_{i0}}/q_i B_0$ and $L_p=2.4$ m approximates the vertical height of the SOL. The velocity-space grid has extents $-4v_{th,s0}\leq v_\parallel \leq 4 v_{th,s0}$ and $0\leq\mu\leq6T_{s0}/B_0$, where $v_{th,s0}=\sqrt{T_{s0}/m_s}$ and $B_0=B_\text{axis}R_0/R_c$. The low-$\beta$ simulations presented here use piecewise-linear ($p=1$) basis functions, with $(N_x,N_y,N_z,N_{v_\parallel},N_\mu)=(32,64,16,10,5)$ the number of cells in each dimension. At high-$\beta$, due to the increased computational cost, the resolution is lowered to $(N_x,N_y,N_z,N_{v_\parallel},N_\mu)=(16,32,14,10,5)$. For $p=1$, one should double these numbers to obtain the equivalent number of grid points for comparison with standard grid-based gyrokinetic codes.

The radial boundary conditions model conducting walls at the radial ends of the domain, given by the Dirichlet boundary condition $\phi=A_\parallel=0$. The condition $\phi=0$ effectively prevents ${\bf E \times B}$ flows into walls, while $A_\parallel=0$ makes it so that (perturbed) field lines never intersect the walls. For the latter, one can think of image currents in the conducting wall that mirror currents in the domain, resulting in exact cancellation of the perpendicular magnetic fluctuations at the wall. Also, the magnetic curvature and $\nabla B$ drifts do not have a radial component in this helical magnetic geometry. These boundary conditions on the fields are thus sufficient to guarantee that there is no flux of the distribution function into the radial walls. Conducting-sheath boundary conditions are applied in the $z$-direction \cite{shi2017gyrokinetic,shi2019full} to model the Debye sheath (the dynamics of which is beyond the gyrokinetic ordering), which partially reflects one species (typically electrons) and fully absorbs the other species depending on the sign of the sheath potential. This involves solving the gyrokinetic Poisson equation for $\phi$ at the $z$-boundary (i.e. sheath entrance), and using the resulting sheath potential to determine a cutoff velocity below which particles (typically low energy electrons) are reflected by the sheath. {Notably, this boundary condition allows current fluctuations in and out of the sheath. This differs from the standard logical sheath boundary condition \cite{parker1993suitable} which imposes zero net current to the sheath by assuming ion and electron currents at the sheath entrance are equal at all times.} The fields do not require a boundary condition in the $z$-direction since only perpendicular derivatives appear in the field equations. The simulations are carried out in a sheath-limited regime but there can be electrical disconnection from the plasma sheath if the Alfv\'en speed is slow enough. Periodic boundary conditions are used in the $y$-direction. The simulation parameters roughly approximate an H-mode deuterium plasma in the NSTX SOL: $B_\text{axis}=0.5$ T, $R_0=0.85$ m, $a=0.5$ m, $T_{e0}=T_{i0}=40$ eV. To model particles and heat from the core crossing the separatrix, we apply a non-drifting Maxwellian source of ions and electrons,
\begin{equation}
    S_{s} = \frac{n_S(x,z)}{(2\pi T_S/m_{s})^{3/2}}\exp\left(-\frac{m_{s} v^2}{2 T_S}\right),
\end{equation}
with source temperature $T_{S}=70$ eV for both species and $v^2 = v_\parallel^2 + 2 \mu B/m_s$. The source density is given by
\begin{equation}
    n_S(x,z) = \begin{cases}
    S_0\exp\left(\frac{-(x-x_S)^2}{(2\lambda_S)^2}\right)\qquad \qquad &|z|<L_z/4\\
    0 \qquad &\mathrm{otherwise}
    \end{cases}
\end{equation}
so that $x_S-3\lambda_S < x < x_S+3\lambda_S$ delimits the source region, and we take $x_S=1.3$ m and $\lambda_S=0.005$ m. This localized particle source structure in $z$-space results in plasma ballooning out largely in the centre of the magnetic field line. The source particle rate $S_0$ is chosen so that the total (ion plus electron) source power matches the desired power into the simulation domain, $P_\mathrm{src}$. Since we only simulate a flux-tube-like fraction of the whole SOL domain, $P_\mathrm{src}$ is related to the total SOL power, $P_\mathrm{SOL}$, by $P_\mathrm{src} = P_\mathrm{SOL}L_y L_z/(2\pi R_c L_\mathrm{pol})\approx0.115 P_\mathrm{SOL}$. Amplitudes are adjusted to approximate $P_\mathrm{SOL}=5.4$ MW crossing into these open flux surfaces at low-$\beta$ conditions relevant to NSTX \cite{Zweben_2015}. An artificially elevated density case with $P_\mathrm{SOL}=54.0$ MW is also tested to study edge turbulence at high-$\beta$. The collision frequency is comparable in magnitude to the inverse autocorrelation time of electron density fluctuations at low-$\beta$. For the high-$\beta$ case, $\nu$ is found to be about $10 \times$ larger, and the plasma thus sits in a strongly collisional regime. Simulations reach a quasi-steady state with the sources balanced by end losses to the sheath, although there is no neutral recycling included yet which is a focus of ongoing work \cite{bernard2020a}. Unlike in \citet{shi2019full}, no numerical heating nor source floors are applied in the algorithm to ensure positivity.

In summary, the full-$f$ nonlinear electromagnetic gyrokinetic turbulence simulations of the NSTX plasma boundary region employ the lowest-order, i.e. guiding-center or drift-kinetic limit, of the system. Implementing gyroaveraging effects given by the next order terms in advanced geometries is the focus of future work. These present approximations in modern full-$f$ global gyrokinetic simulations should be kept in mind when attributing any similarities or differences to two-fluid theory in the next sections since the gyrokinetic formulation can itself be improved. The turbulence simulations presented are thus a reflection of the current forefront of numerical modelling. A full exposition of the derivation and benchmarking including the energy-conserving discontinuous Galerkin scheme applied in \texttt{Gkeyll} for the discretization of the gyrokinetic system in 5-dimensional phase space along with explicit time-stepping and avoidance of the Amp\`ere cancellation problem can be found in \citet{mandell_thesis}. 

\section{\label{sec:level3}Machine learning fluid theory}

A fundamental goal in computational plasma physics is determining the minimal complexity necessary in a theory to sufficiently represent (anomalous) observations of interest for predictive fusion reactor simulations. Convection-diffusion equations with effective mean transport coefficients are widely utilized \cite{Reiter_1991,SOLPS_DEKEYSER2019125} but insufficient in capturing edge plasma turbulence where scale separation between the equilibrium and fluctuations is not justified \cite{NAULIN2007,mandell_thesis}. Other reduced turbulence models such as classical magnetohydrodynamics are unable to resolve electron and ion temperature dynamics with differing energy transport mechanisms in the edge \cite{TeTi_2008,TeTi_2009,TeTi_2016}. Following the framework set forth in \citet{mathews2020uncovering}, we consider here the widely used first-principles-based two-fluid drift-reduced Braginskii equations \cite{Braginskii1965RvPP,francisquez_thesis} in the electrostatic limit relevant to low-$\beta$ conditions in the edge of fusion experiments \cite{Zweben_2015} for comparison to gyrokinetic modelling \cite{mandell_hakim_hammett_francisquez_2020}. This is also a full-$f$ \cite{Belli_2008,Full-F,Held_2020} nonlinear model in the sense that the evolution of the equilibrium and fluctuating components of the solution are merged and relative perturbation amplitudes can be of order unity but evaluated in the fluid limit. This two-fluid theory assumes the plasma is magnetized ($\Omega_i \gg \frac{\partial}{\partial t}$), collisional ($\nu_{ei} \gg \frac{\partial}{\partial t}$), and quasineutral ($\nabla \cdot \v{j} \approx 0$) with the reduced perpendicular fluid velocity given by ${\bf E \times B}$, diamagnetic, and ion polarization drifts. After neglecting collisional drifts and terms of order $m_e/m_i$, one arrives at the following set of equations (in Gaussian units) governing the plasma's density ($n \approx n_e \approx n_i$), vorticity ($\gvort$), parallel electron velocity ($\vpe$), parallel ion velocity ($\vpi$), electron temperature ($T_e$), and ion temperature ($T_i$) \cite{francisquez2020fluid,francisquez_thesis}

\begin{eqnal}\label{eq:nDotGDBH}
\d{^e n}{t} = -\frac{2c}{B}\left[n\curv{\phi}-\frac{1}{e}\curv{p_e}\right] -n\delpar{\vpe} +\nSrcN+\mathcal{D}_{n}
\end{eqnal}
\begin{eqnal}\label{eq:wDotGDBH}
\pd{\gvort}{t} &= \frac{2c}{eB}\left[\curv{p_e}+\curv{p_i}\right]-\frac{1}{em_i \Omega_i}\curv{G_i} \\
&\quad+\frac{1}{e}\delpar{\cur}-\div{\left\lbrace\frac{nc^2}{\Omega_i B^2}\left[\phi,\gradperp{\phi}+\frac{\gradperp{p_i}}{en}\right]\right. \\ 
&\quad\left.+\frac{nc}{\Omega_i B}\vpi\delpar{\left(\gradperp{\phi}+\frac{\gradperp{ p_i}}{en}\right)}\right\rbrace}+\mathcal{D}_{\gvort}
\end{eqnal}
\begin{eqnal}\label{eq:vpeDotGDBH}
\d{^e\vpe}{t} &= \frac{1}{m_e}\left(e\delpar{\phi}-\frac{\delpar{p_e}}{n}-0.71\delpar{T_e} + e\eta_\parallel\cur \right) \\
&\quad + \frac{2}{3} \frac{\delpar{G_e}}{n} + \frac{2cT_e}{eB}\curv{\vpe}+\momSrce+\mathcal{D}_{\vpe}
\end{eqnal}
\begin{eqnal}\label{eq:vpiDotGDBH}
\d{^i\vpi}{t} &= \frac{1}{m_i}\left(-e\delpar{\phi}-\frac{\delpar{p_i}}{n}+0.71\delpar{T_e} - e\eta_\parallel\cur \right)\\
&+\frac{2 T_e}{3n}\frac{\delpar{G_i}}{n}-\frac{2cT_i}{eB}\curv{\vpi}+\momSrci+\mathcal{D}_{\vpi}
\end{eqnal}
\begin{eqnal}\label{eq:TeDotGDBH}
\d{^e T_e}{t} = \frac{2T_e}{3n}\left[\d{^e n}{t} + \frac{1}{T_e}\delpar \kappa^e_\parallel \delpar T_e + \frac{5n}{m_e \Omega_e} \curv{T_e} \right.\\ \left. + \eta_\parallel \frac{\cur^2}{T_e} + \frac{0.71}{e}(\delpar{\cur} - \frac{\cur}{T_e}\delpar{T_e}) + \frac{1}{T_e} \enerSrceN \right] + \mathcal{D}_{T_e},
\end{eqnal}
\begin{eqnal}\label{eq:TiDotGDBH}
\d{^i T_i}{t} &= \frac{2T_i}{3n}\left[\d{^i n}{t} + \frac{1}{T_i}\delpar \kappa^i_\parallel \delpar T_i \right.\\
&\quad \left. - \frac{5n}{m_i \Omega_i} \curv{T_i} + \frac{1}{T_i} \enerSrciN \right] + \mathcal{D}_{T_i}
\end{eqnal}
whereby the field-aligned electric current density is $\cur = en\left(\vpi - \vpe\right)$, the stress tensor's gyroviscous terms with some FLR effects contain $G_s = \eta^s_0 \left\lbrace 2\delpar{v_{\parallel s}}+c\left[\curv{\phi} + \curv{p_s}/(q_s n)\right]\right\rbrace$, and $\eta^s_0$ is the viscosity. The convective derivatives are $d^s f/dt = \partial_t f + (c/B)\left[\phi,f\right] + v_{\parallel s}\delpar{f}$ with $\left[F,G\right] = \bhatZ \times \nabla F \cdot \nabla G$ and $\bhatZ$ representing a constant unit vector parallel to the unperturbed magnetic field. Consistent with the gyrokinetic modelling, the field's magnitude, $B$, decreases over the major radius of the torus ($B\propto1/R$), and $\gv{\kappa} = -{\bf{\hat{R}}}/R$. Curvature drifts are encoded via diamagnetism in this fluid representation \cite{fluidvparticle_drifts} with the curvature operator, $\curv{f} = \bhatZ \times \gv{\kappa} \cdot \grad{f}$, following \eqref{eq:GkCurv} but with an unperturbed field vector distinguishing it from electromagnetic theory. The constant coefficients $\kappa^s_\parallel$ and $\eta^s_\parallel$ correspond to parallel thermal conductivity and electrical resistivity, respectively. The 3-dimensional shearless field-aligned coordinate system over which the drift-reduced Braginskii theory is formulated in the physics-informed machine learning framework exactly matches the gyrokinetic code's geometry. The two-fluid model similarly consists of deuterium ions and electrons with real masses (i.e. $m_i = 3.34 \times 10^{-27} \text{ kg}$ and $m_e = 9.11\times 10^{-31} \text{ kg}$).

To consistently calculate the electric field response and integrate \eqref{eq:nDotGDBH}--\eqref{eq:TiDotGDBH} in time, classically one would compute the turbulent $\phi$ in drift-reduced Braginskii theory by directly invoking quasineutrality and numerically solving the boundary value problem given by the divergence-free condition of the electric current density \cite{DRB_consistency3,Zholobenko_2021}. For the purposes of comparison with gyrokinetic modelling, we apply a novel technique for computing the turbulent electric field consistent with the drift-reduced Braginskii equations without explicitly evaluating $\nabla \cdot \v{j} = 0$ nor directly applying the Boussinesq approximation. Namely, this work applies a validated physics-informed deep learning framework \cite{mathews2020uncovering} to infer the gauge-invariant $\phi$ directly from \eqref{eq:nDotGDBH} and \eqref{eq:TeDotGDBH} for direct analysis of electron pressure and electric field fluctuations in nonlinear global electromagnetic gyrokinetic simulations and electrostatic two-fluid theory. To review the pertinent aspects of the machine learning methodology \cite{mathews2020uncovering}, every dynamical variable in equations \eqref{eq:nDotGDBH}--\eqref{eq:TiDotGDBH} is approximated by its own fully-connected neural network with local spatiotemporal points $(x,y,t)$ from a reduced 2-dimensional domain as the only inputs to the initial layer. The dynamical variable being represented by each network is the sole output. In the middle of the architecture, every network consists of 8 hidden layers with 50 neurons per hidden layer and hyperbolic tangent activation functions all utilizing Xavier initialization \cite{GlorotAISTATS2010}. The turbulent electric field consistent with drift-reduced Braginskii theory is then learnt by optimizing the networks against both $n_e$ and $T_e$ measurements along with the physical theory encoded by \eqref{eq:nDotGDBH} and \eqref{eq:TeDotGDBH}. To be precise, as visualized in Figure 3 of \citet{mathews2020uncovering}, learning proceeds by training the $n_e$ and $T_e$ networks against the average $\mathcal{L}_2$-norm of their respective errors
\begin{equation}\label{eq:L_n_DotGDBH}
    \mathcal{L}_{n_e} = \frac{1}{N_0}\sum_{i=1}^{N_0} \lvert n^*_{e}(x^i_0,y^i_0,z^i_0,t^i_0) - n^i_{e,0} \rvert^2,
\end{equation}
\begin{equation}\label{eq:L_Te_DotGDBH}
    \mathcal{L}_{T_e} = \frac{1}{N_0}\sum_{i=1}^{N_0} \lvert T^*_{e}(x^i_0,y^i_0,z^i_0,t^i_0) - T^i_{e,0} \rvert^2,
\end{equation}
where $\lbrace x_0^i,y_0^i,z_0^i,t_0^i,n_{e,0}^i,T_{e,0}^i\rbrace^{N_0}_{i=1}$ correspond to the set of observed 2-dimensional data and the variables $n^*_e$ and $T^*_e$ symbolize the electron density and temperature predicted by the networks, respectively. Properties of all other dynamical variables in the 6-field turbulence model are unknown. The only existing requirement on the observational data in this deep learning framework is that the measurements' spatial and temporal resolutions should be be finer than the autocorrelation length and time, respectively, of the turbulence structures being analyzed. This scale condition is well-satisfied for the low-$\beta$ case and marginally-satisfied for the high-$\beta$ data analyzed. Model loss functions are synchronously learnt by recasting \eqref{eq:nDotGDBH} and \eqref{eq:TeDotGDBH} in the following implicit form

\begin{eqnal}\label{eq:f_n_DotGDBH}
f_{n_e} &\coloneqq -\d{^e n}{t} -\frac{2c}{B}\left[n\curv{\phi}-\frac{1}{e}\curv{p_e}\right]\\ &\quad-n\delpar{\vpe} +\nSrcN+\mathcal{D}_{n},
\end{eqnal}

\begin{eqnal}\label{eq:f_Te_DotGDBH}
f_{T_e} &\coloneqq -\d{^e T_e}{t} + \frac{2T_e}{3n}\left[\d{^e n}{t} + \frac{1}{T_e}\delpar \kappa^e_\parallel \delpar T_e \right.\\& \left. + \frac{5n}{m_e \Omega_e} \curv{T_e} + \eta_\parallel \frac{\cur^2}{T_e} + \frac{0.71}{e}(\delpar{\cur} \right.\\ & \left. - \frac{\cur}{T_e}\delpar{T_e}) + \frac{1}{T_e} \enerSrceN \right] + \mathcal{D}_{T_e},
\end{eqnal}
and then further normalized into dimensionless form \cite{francisquez2020fluid,mathews2020uncovering}. These unitless evolution equations of $n_e$ and $T_e$ from the two-fluid theory are jointly optimized using
\begin{equation}\label{eq:L_f_n_DotGDBH}
    \mathcal{L}_{f_{n_e}} = \frac{1}{N_f}\sum_{i=1}^{N_f} \lvert f^{*}_{n_e}(x^i_f,y^i_f,z^i_f,t^i_f) \rvert^2,
\end{equation}
\begin{equation}\label{eq:L_f_Te_DotGDBH}
    \mathcal{L}_{f_{T_e}} = \frac{1}{N_f}\sum_{i=1}^{N_f} \lvert f^{*}_{T_e}(x^i_f,y^i_f,z^i_f,t^i_f) \rvert^2,
\end{equation}
where $\lbrace x_f^i,y_f^i,z_f^i,t_f^i\rbrace^{N_f}_{i=1}$ denote the set of collocation points, and $f^{*}_{n_e}$ and $f^{*}_{T_e}$ are the null partial differential equations prescribed by \eqref{eq:f_n_DotGDBH} and \eqref{eq:f_Te_DotGDBH} in normalized form directly evaluated by the neural networks. The set of collocation points over which the partial differential equations are evaluated correspond to the positions of the observed electron pressure data, i.e. $\lbrace x_0^i,y_0^i,z_0^i,t_0^i\rbrace^{N_0}_{i=1} = \lbrace x_f^i,y_f^i,z_f^i,t_f^i\rbrace^{N_f}_{i=1}$. This physics-informed framework is consequently a truly deep approximation of drift-reduced Braginskii theory since every dynamical variable is represented by its own fully-connected deep network which is then further differentiated to build the cumulative computation graph reflecting every single term in the evolution equations. Loss functions are optimized with mini-batch sampling where $N_0 = N_f = 1000$ using L-BFGS---a quasi-Newton optimization algorithm \cite{10.5555/3112655.3112866}---for 20 hours over 32 cores on Intel Haswell-EP processors. We highlight that the only locally observed dynamical quantities in these equations are 2-dimensional views of $n_e$ and $T_e$ (to emulate gas puff imaging \cite{Mathews2021}) without any explicit information about boundary conditions nor initializations nor ion temperature dynamics nor parallel flows. This machine learning framework's collocation grid consists of a continuous spatiotemporal domain without time-stepping nor finite difference schema in contrast with standard numerical codes. All analytic terms encoded in these equations including high-order operators are computed exactly by the neural networks without discretization. In that sense, it is a potentially higher fidelity continuous representation of the continuum equations. While the linearized polarization density---analogous to the Boussinesq approximation---is employed in the gyrokinetic simulations, no such approximations are explicitly applied by our physics-informed neural networks.

With sparse availability of measurements in fusion experiments, designing diagnostic techniques for validating turbulence theories with limited information is crucial. On this point, we note that our framework can potentially be adapted to experimental measurements of electron pressure \cite{griener_continuous_2020,Bowman_2020,Mathews2020,Mathews2021}. To handle the particular case of 2-dimensional turbulence data, we essentially assume slow variation of dynamics in the $z$-coordinate and effectively set all parallel derivatives to zero. In computational theory comparisons where no such limitations exist, training on $n_e$ and $T_e$ in 3-dimensional space over long macroscopic timescales can be easily performed via segmentation of the domain and parallelization, but a limited spatial view away from numerical boundaries with reduced dimensionality is taken to mirror experimental conditions for field-aligned observations \cite{Zweben_2017} and fundamentally test what information is indispensable to compare kinetic and fluid turbulence. To compare the gyrokinetic and two-fluid theories as directly as possible, we analyze the toroidal simulations at $z = L_z/3$ where no applied sources are present. We further remark that, beyond the inclusion of appropriate collisional drifts and sources, this technique is quite robust and generalizable to boundary plasmas with multiple ions and impurities present due to the quasi-neutrality assumptions underlying the two-fluid theory \cite{multi_species}. This turbulence diagnostic analysis technique is hence easily transferable, which permits its systematic application across magnetic confinement fusion experiments where the underlying physical model governing turbulent transport is consistent. The framework sketched can thus be extended to different settings especially in the interdisciplinary study (both numerical and experimental) of magnetized collisional plasmas in space propulsion systems and fusion energy and astrophysical environments if sufficient observations exist. A full description and testing of the overall deep learning technique is provided in \citet{mathews2020uncovering}.
 
\section{\label{sec:level4}Quantifying plasma turbulence consistency}

\begin{figure*}[ht]
\includegraphics[width=1.0\linewidth]{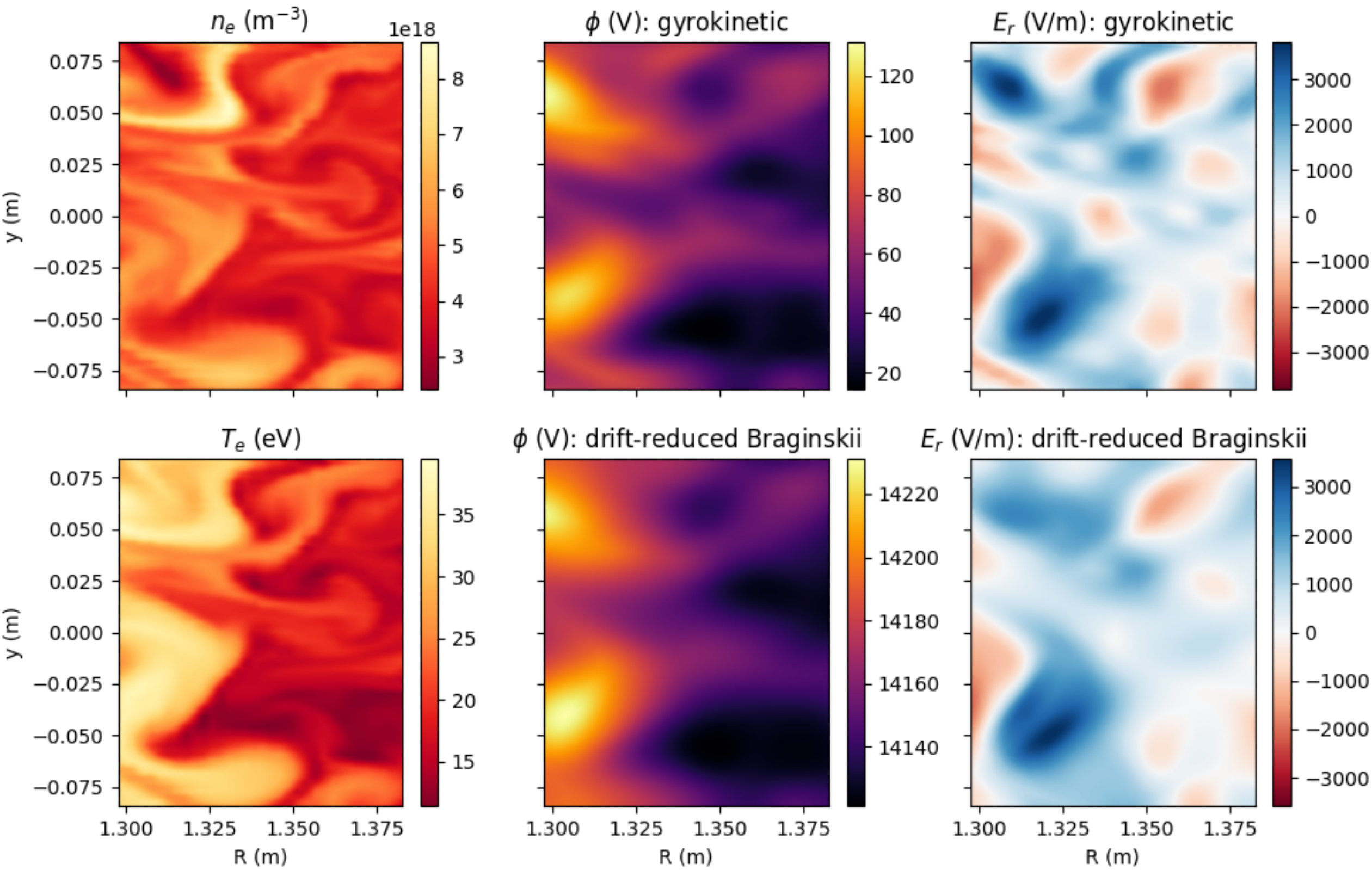}
\caption{\label{PlasmaPINN_lowbeta}The turbulent electric potential, $\phi$ (a gauge-invariant quantity which is equivalent up to a scalar constant offset), and radial electric field, $E_r$, concomitant with electron pressure fluctuations as predicted by electrostatic drift-reduced Braginskii theory and electromagnetic gyrokinetic modelling in low-$\beta$ conditions are in good quantitative agreement. The two-fluid theory's $\phi$ and $E_r$ are based upon the training the physics-informed deep learning framework while the gyrokinetic results are from the discontinuous Galerkin numerical solver.}
\end{figure*}

\begin{figure}[ht]
\includegraphics[width=0.89\linewidth]{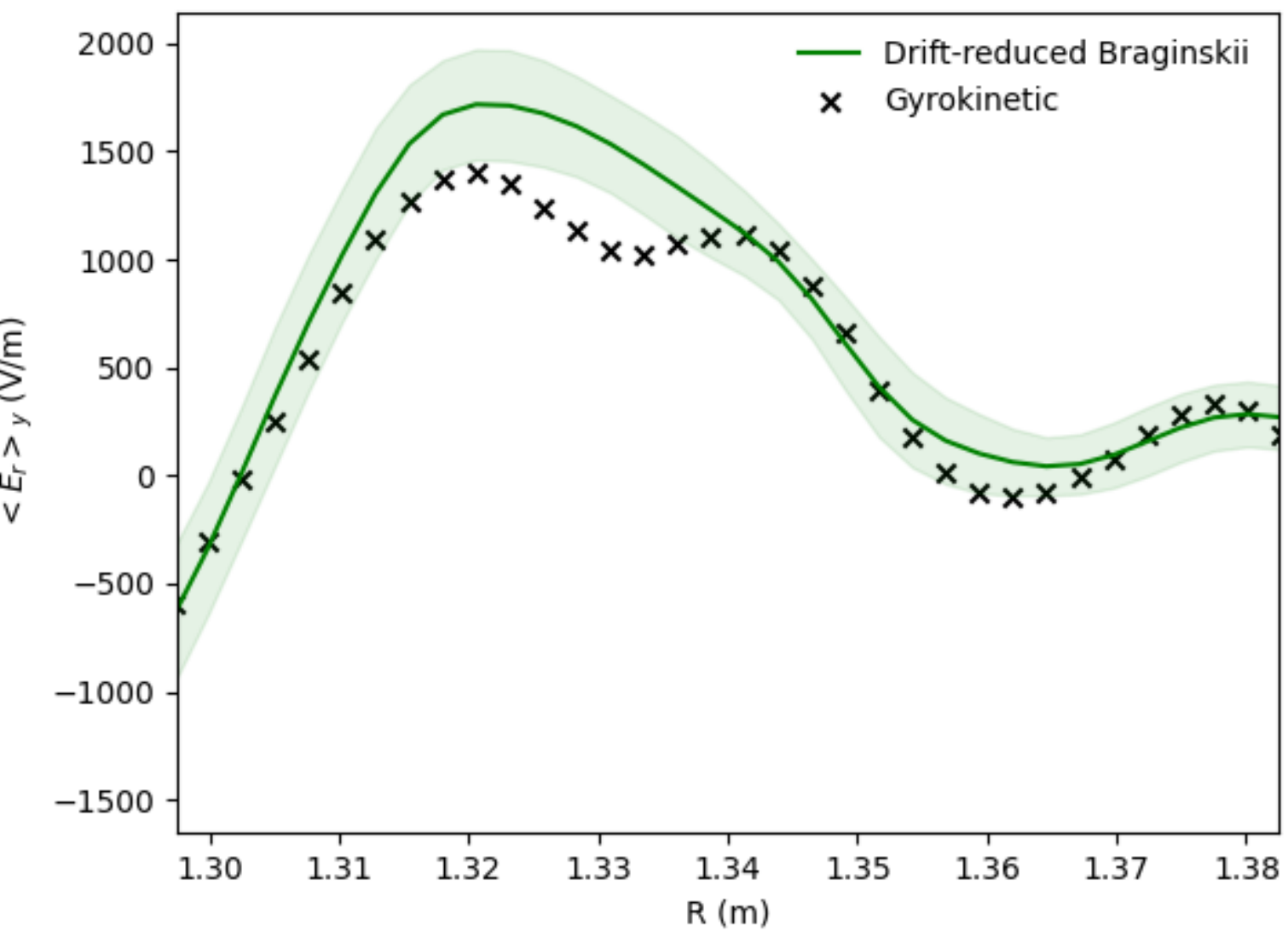}
\caption{\label{PlasmaPINN_lowbeta_avg}The $y$-averaged turbulent radial electric field, $\langle E_r\rangle_y$, as predicted by electrostatic drift-reduced Braginskii theory and electromagnetic gyrokinetic modelling at low-$\beta$. The results plotted for the drift-reduced Braginskii output here are based upon collecting 100 independently-trained physics-informed neural networks.}
\end{figure}

A defining characteristic of a nonlinear theory is how it mathematically connects dynamical variables. The focus of this work is quantitatively examining the nonlinear relationship extant in plasma turbulence models between electron pressure and electric field fluctuations. As outlined in Section \ref{sec:level3}, using a custom physics-informed deep learning framework whereby the drift-reduced Braginskii equations are embedded in the neural networks via constraints in the form of implicit partial differential equations, we demonstrate for the first time direct comparisons of instantaneous turbulent fields between electrostatic two-fluid theory and electromagnetic long-wavelength gyrokinetic modelling in low-$\beta$ helical plasmas with results visualized in Figure \ref{PlasmaPINN_lowbeta}. This multi-network physics-informed deep learning framework enables direct comparison of drift-reduced Braginskii theory with gyrokinetics and bypasses demanding limitations in standard forward modelling numerical codes which must precisely align initial conditions, physical sources, numerical diffusion, and boundary constraints (e.g. particle and heat fluxes into walls) in both gyrokinetic and fluid representations when classically attempting comparisons of turbulence simulations and statistics. Further, the theoretical and numerical conservation properties of these simulations ordinarily need to be evaluated which can be a significant challenge especially when employing disparate numerical methods with differing discretization schema and integration accuracy altogether \cite{francisquez2020fluid}. We are able to overcome these hurdles by training on partial electron pressure observations simultaneously with the plasma theory sought for comparison. We specifically prove that the turbulent electric field predicted by drift-reduced Braginskii two-fluid theory is largely consistent with long-wavelength gyrokinetic modelling in low-$\beta$ helical plasmas. This is also evident if analyzing $y$-averaged radial electric field fluctuations and accounting for the inherent scatter ($\sigma_{PINN}$) from the stochastic optimization employed as displayed in Figure \ref{PlasmaPINN_lowbeta_avg}. To clarify the origins of $\sigma_{PINN}$, every time our physics-informed deep learning framework is trained from scratch against the electron pressure observations, the learned turbulent electric field will be slightly different due to the random initialization of weights and biases for the networks along with mini-batch sampling during training. To account for this stochasticity, we have trained our framework anew 100 times while collecting the predicted turbulent $E_r$ after each optimization. the quantity $\sigma_{PINN}$ corresponds to the standard deviation from this collection. The two-fluid model's results displayed in Figure \ref{PlasmaPINN_lowbeta_avg} is thus based upon computing $\langle E_r\rangle_y$ from 100 independently-trained physics-informed machine learning frameworks. Their turbulent outputs are averaged together to produce the mean, while the scatter associated with the 100 realizations---which approximately follows a normal distribution---comprises the shaded uncertainty interval spanning 2 standard deviations. As visualized, the $\langle E_r\rangle_y$ profiles predicted by the the electromagnetic gyrokinetic simulation and electrostatic drift-reduced Braginskii model are generally in agreement within error bounds at low-$\beta$.

\begin{figure*}[ht]
\includegraphics[width=1.0\linewidth]{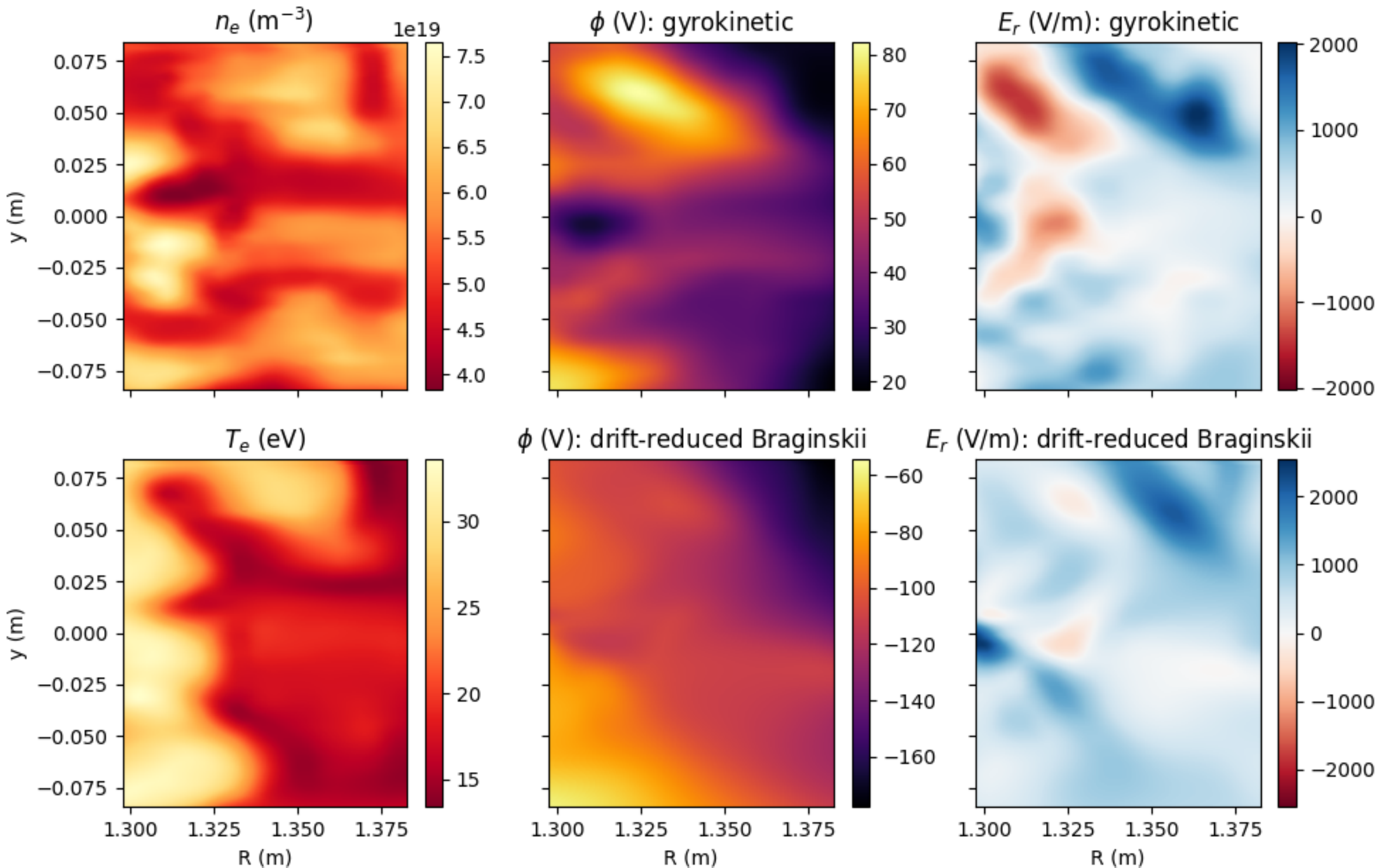}
\caption{\label{PlasmaPINN_highbeta}The turbulent electric potential, $\phi$ (a gauge-invariant quantity which is equivalent up to a scalar constant offset), and radial electric field, $E_r$, concomitant with electron pressure fluctuations as predicted by electrostatic drift-reduced Braginskii theory and electromagnetic gyrokinetic modelling in high-$\beta$ conditions are quantitatively inconsistent. The two-fluid theory's $\phi$ and $E_r$ are based upon the training of the physics-informed deep learning framework while the gyrokinetic results are from the discontinuous Galerkin numerical solver.}
\end{figure*}
\begin{figure}[ht]
\includegraphics[width=0.875\linewidth]{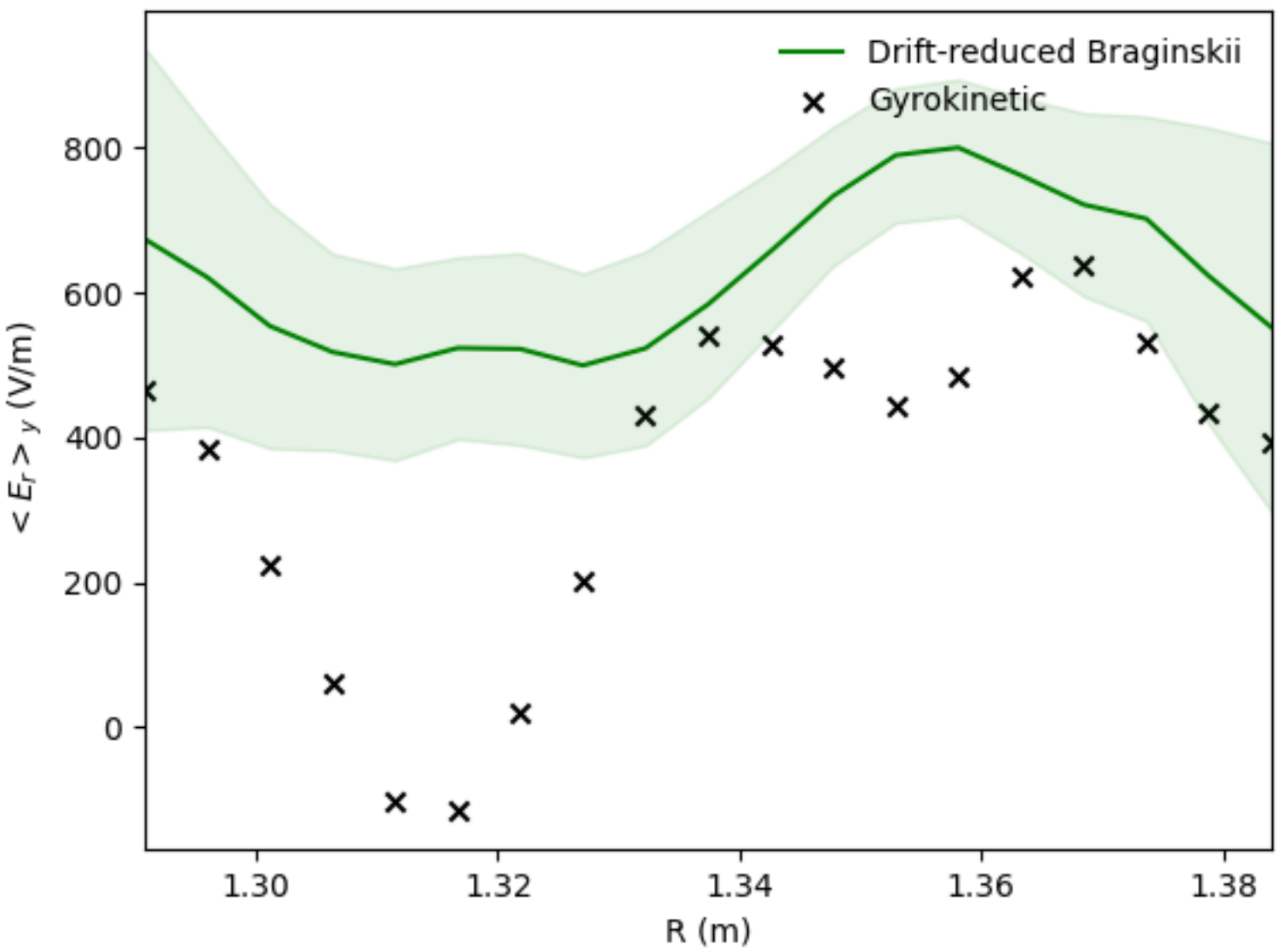}
\caption{\label{PlasmaPINN_highbeta_avg}The $y$-averaged turbulent radial electric field, $\langle E_r\rangle_y$, as predicted by electrostatic drift-reduced Braginskii theory and electromagnetic gyrokinetic modelling at high-$\beta$. The results plotted for the drift-reduced Braginskii output here are based upon collecting 100 independently-trained physics-informed neural networks.}
\end{figure}

\begin{figure}
\includegraphics[width=1.0\linewidth]{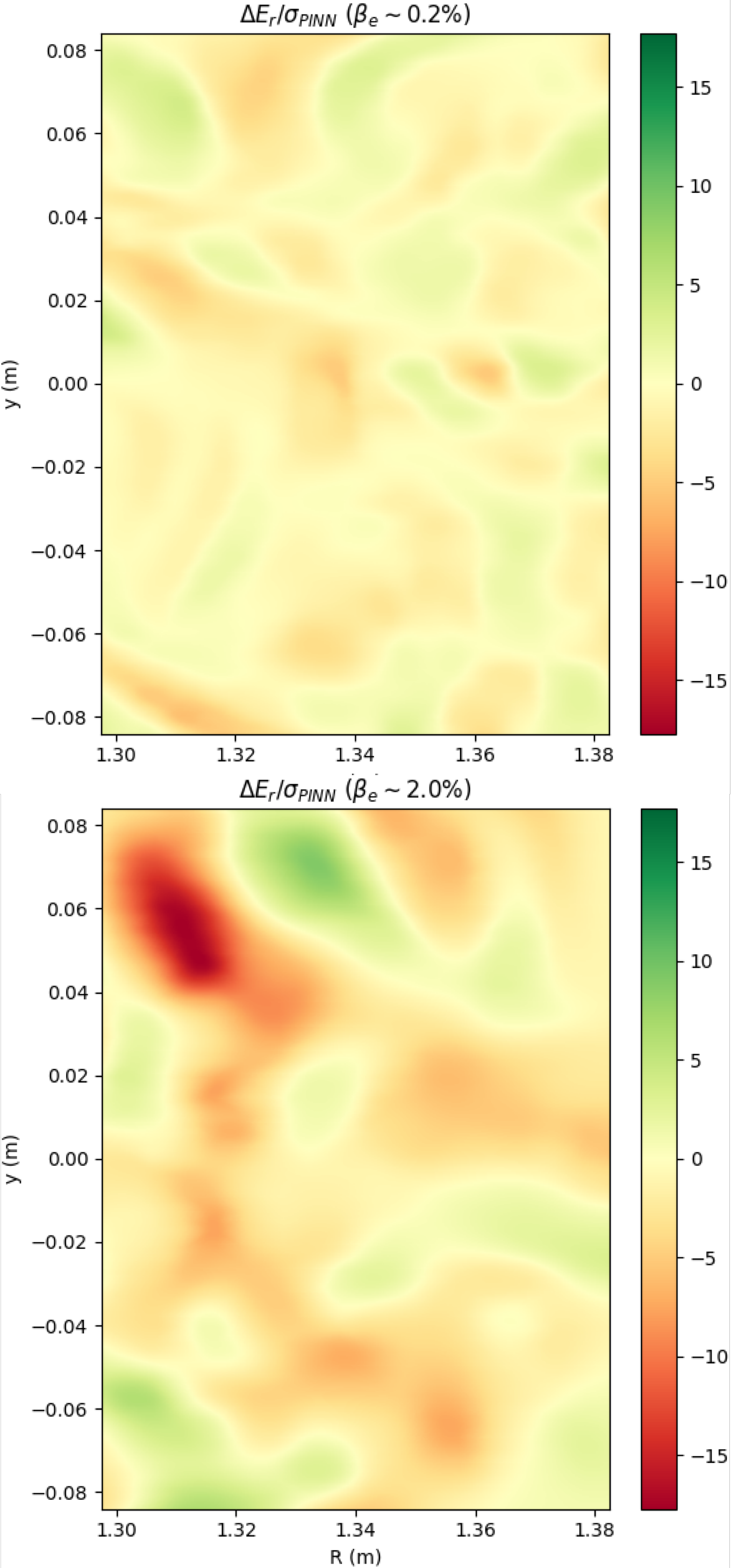}
\caption{\label{PlasmaPINN_inst_err}The relative error in the instantaneous radial electric field fluctuations ($\Delta E_r / \sigma_{PINN}$) between electrostatic drift-reduced Braginskii theory and electromagnetic long-wavelength gyrokinetic modelling is displayed in low-$\beta$ (top) and high-$\beta$ (bottom) conditions. While all errors in the low-$\beta$ scenario are generally within approximately 3--4 standard deviations and representative of mostly good quantitative agreement, one must go to over $15\sigma_{PINN}$ to fully account for the turbulent fields using an electrostatic two-fluid theory at $\beta_e \sim 2\%$. The results are based upon collecting 100 independently-trained physics-informed neural networks to compute the turbulent $E_r$ and the intrinsic scatter in these predictions.}
\end{figure}

In high-$\beta$ conditions where the particle source is artificially increased by 10$\times$, electromagnetic effects become important and the electrostatic two-fluid theory is markedly errant juxtaposed with electromagnetic gyrokinetic simulations when considering the comparison in Figure \ref{PlasmaPINN_highbeta}. As remarked above, multiple realizations are conducted to analyze the sample statistics of the learned turbulent fields consistent with drift-reduced Braginskii two-fluid theory based solely upon the intrinsic scatter during training to account for the stochastic nature of the optimization. By collecting 100 independently-trained realizations, the inherent uncertainty linked to this intrinsic scatter can be evaluated as demonstrated in Figure \ref{PlasmaPINN_highbeta_avg}. These discrepancies indicate that the fluctuations in $\bf B_\perp$, which are evaluated by solving the parallel Amp\`ere equation (or, equivalently, generalized Ohm's law), cannot be neglected when considering plasma transport across the inhomogeneous background magnetic field as in electrostatic theory. While the fluid approximation is generally expected to be increasingly accurate at high density due to strong coupling between electrons and ions, these results underline the importance of electromagnetic effects even in shear-free high-$\beta$ plasmas as found in planetary magnetospheres and fusion experiments (e.g. dipole confinement \cite{LDX-experiment}), and for the first time enables the degree of error between instantaneous fluctuations to be precisely quantified across turbulence models. Uncertainty estimates stemming from the stochastic framework in both regimes are reflected in Figure \ref{PlasmaPINN_inst_err}. 

One should note that there are novel and different levels of errors to be mindful of in this evaluation. For example, poor convergence arising from nonuniqueness of the turbulent $E_r$ found during optimization against the drift-reduced Braginskii equations, or $\mathcal{L}_{f_{n_e}}$ and $\mathcal{L}_{f_{T_e}}$ remaining non-zero (and not below machine precision) even after training \cite{mathews2020uncovering}. These potential errors exist on top of standard approximations in the discontinuous Galerkin numerical scheme representing the underlying gyrokinetic theory such as the implemented Dougherty collision operator. Notwithstanding, when comparing the electrostatic drift-reduced Braginskii theory to electromagnetic long-wavelength gyrokinetic simulations at low-$\beta$, the results represent good consistency in the turbulent electric field and all observed discrepancies are mostly within the stochastic optimization's expected underlying scatter. Alternatively, when analyzing high-$\beta$ conditions, we observe that the electrostatic two-fluid model cannot explain the turbulent electric field in the electromagnetic gyrokinetic simulations. In particular, $\Delta E_r \gtrsim 15\sigma_{PINN}$ in the bottom plot of Figure \ref{PlasmaPINN_inst_err}, where $\Delta E_r$ is the difference in the instantaneous $E_r$ predicted by two-fluid theory and gyrokinetics. This signals that the two models' turbulent $E_r$ fluctuations are incompatible at high-$\beta$ conditions while quantifying the separation. 

Our multi-network physics-informed deep learning framework demonstrates the suitability of electrostatic two-fluid theory as a good approximation of turbulent electric fields in modern gyrokinetic simulations for low-$\beta$ helical plasmas with sufficient initial and boundary conditions. Conversely, the electrostatic turbulence model is demonstrably insufficient for high-$\beta$ plasmas. This finding is indicative of the importance of including electromagnetic effects such as magnetic flutter in determining cross-field transport even at $\beta_e \sim 2\%$. But field line perturbations and the reduced numerical representation of gyrokinetic theory are not the only effects at play causing mismatch at high-$\beta$: due to the artificial nature of the strong localized particle source in $z$-space to produce high-$\beta$ conditions, parallel dynamics including electron flows along field lines and Ohmic heating effects become increasingly important. These variables should now be observed or learnt first from the 2-dimensional electron pressure measurements to accurately reconstruct the turbulent electric field which signifies a departure from the expected low-$\beta$ edge of tokamaks. Going forward, consideration of advanced magnetic geometries with squeezing and shearing of fusion plasmas near null-points, which may even couple-decouple upstream turbulence in tokamaks, will be important in the global validation of reduced turbulence models with realistic shaping \cite{Ryutov_geometry,Kuang2018APS,kuang_thesis,Nespoli_2020,Myra_2020}. Also, while there is generally good convergence in the low-$\beta$ gyrokinetic simulations at parameters relevant to the edge of NSTX, numerical convergence in the artificially elevated high-$\beta$ case is currently questionable and a potential source of discrepancy. Running the high-$\beta$ gyrokinetic simulation with proper collision frequency at an improved resolution of $(N_x,N_y,N_z,N_{v_\parallel},N_\mu)=(48,96,18,10,5)$ would cost roughly 4 million CPU-hours to check and we must leave such investigations for future analysis.

As for distinctiveness, we point out that the techniques used for computing the displayed turbulent electric fields in the two cases are markedly different. In particular, the long-wavelength gyrokinetic Poisson equation, which is originally derived from the divergence-free condition of the electric current density, is employed in the gyrokinetic simulations. In contrast, simply the electron fluid evolution equations are used to infer the unknown turbulent field fluctuations consistent with drift-reduced Braginskii theory \cite{mathews2020uncovering}. A principle underlying these models is quasineutrality, but this condition is not sufficient on its own. If one were to apply equilibrium models such as the Boltzmann relation or simple ion pressure balance as expected neoclassically, the turbulent electric field estimates for these nonequilibrium plasmas with nontrivial cross-field transport would be highly inaccurate \cite{mathews2020uncovering}. Further, no external knowledge of boundary conditions such as sheath effects are explicitly provided to the physics-informed deep learning framework, but this information implicitly propagates from the walls into the observations of electron pressure. This novel approach resultantly permits using limited 2D measurements to compare a global 3D fluid turbulence model directly against 5D gyrokinetics beyond statistical considerations for characterizing non-diffusive intermittent edge transport \cite{NAULIN2007}. All in all, the agreement between drift-reduced Braginskii theory and gyrokinetics supports its usage for predicting turbulent transport in low-$\beta$ shearless toroidal plasmas, but an analysis of all dynamical variables is required for the full validation of drift-reduced Braginskii theory. Alternatively, when 2-dimensional experimental electron pressure measurements are available \cite{Furno2008,Mathews2021}, this technique can be used to infer $E_r$ and the resulting structure of turbulent fluxes heading downstream. And, if independently diagnosed measurements of the laboratory plasma's turbulent $E_r$ exist, the overall turbulence theory can be expressly tested in experiment.

\section{\label{sec:level5}Conclusion}

To probe the fundamental question of how similar two distinct turbulence models truly are, using a novel technique to analyze electron pressure fluctuations, we have directly demonstrated that there is good agreement in the turbulent electric fields predicted by electrostatic drift-reduced Braginskii theory and electromagnetic long-wavelength gyrokinetic simulations in low-$\beta$ helical plasmas. As $\beta_e$ is increased to approximately $2\%$, the 2-dimensional electrostatic nature of the utilized fluid theory becomes insufficient to explain the microinstability-induced particle and heat transport. Overall, by analyzing the interconnection between dynamical variables in these global full-$f$ models, physics-informed deep learning can quantitatively examine this defining nonlinear characteristic of turbulent physics. In particular, we can now unambiguously discern when agreement exists between multi-field turbulence theories and identify disagreement when incompatibilities exist with just 2-dimensional electron pressure measurements. This machine learning tool can therefore act as a necessary condition to diagnose when reduced turbulence models are unsuitable, or, conversely, iteratively construct and test theories till agreement is found with observations.

While we focus on the electric field response to electron pressure in this work, extending the analysis to all dynamical variables (e.g. $T_i, \vpe, \vpi$) for full validation of reduced multi-field turbulence models in a variety of regimes (e.g. collisionality, $\beta$, closed flux surfaces with sheared magnetic field lines) using electromagnetic fluid theory is the subject of future work. Also, since plasma fluctuations can now be directly compared across models, as gyrokinetic codes begin including fully kinetic neutrals, this optimization technique can help validate reduced source models to accurately account for atomic and molecular interactions with plasma turbulence in the edge of fusion reactors \cite{Thrysoe_2018,Thrysoe_2020} since these processes (e.g. ionization, recombination) affect the local electric field. Further progress in the gyrokinetic simulations such as the improved treatment of gyro-averaging effects \cite{brizard2007foundations}, collision operators \cite{francisquez2020conservative}, and advanced geometries \cite{mandell_thesis} will enable better testing and discovery of hybrid reduced theories as well \cite{zhu2021drift}. For example, in diverted reactor configurations, electromagnetic effects become increasingly important for transport near X-points where $\beta_{p} \rightarrow \infty$. A breakdown of Alfv\'{e}n's theorem in these regions can also arise due to the impact of Coulomb collisions and magnetic shear contributing to an enhanced perpendicular resistivity \cite{Myra-X-point} which presents an important test case of non-ideal effects within reduced turbulence models. While our work supports the usage of electrostatic two-fluid modelling, with adequate initial and boundary conditions, over long-wavelength gyrokinetics for low-$\beta$ magnetized plasmas without magnetic shear, a comparison of all dynamical variables beyond the turbulent electric field is required for a full validation of the reduced model. Further investigations into reactor conditions may suggest the modularization of individually validated fluid-kinetic turbulence models across different regions in integrated global device modelling efforts \cite{hakim2019discontinuous,Merlo2021_XGC_GENE}. This task can now be efficiently tackled through pathways in deep learning as demonstrated by this new class of validation techniques. In addition, precisely understanding the fundamental factors--both physical and numerical--determining the prediction interval, $\sigma_{PINN}$, is the subject of ongoing research in analyzing the nature (e.g. uniqueness, smoothness) of chaotic solutions to fluid turbulence equations and the chaotic convergence properties of physics-informed neural networks. 

\begin{acknowledgments}
The authors wish to thank A.Q. Kuang, D.R. Hatch, and G.W. Hammett for insights shared and helpful discussions. Gyrokinetic simulations with the \texttt{Gkeyll} code (\url{https://gkeyll.readthedocs.io/en/latest/}) were performed on the Perseus cluster at Princeton University and the Cori cluster at NERSC. All further analysis and codes run were conducted on MIT's Engaging cluster and we are grateful for assistance with computing resources provided by all teams. The work is supported by the Natural Sciences and Engineering Research Council of Canada (NSERC) by the doctoral postgraduate scholarship (PGS D), Manson Benedict Fellowship, and the U.S. Department of Energy (DOE) Office of Science under the Fusion Energy Sciences program by contracts DE-SC0014264, DE-SC0014664, DE-AC02-09CH11466 via the Scientific Discovery through Advanced Computing (SciDAC) program, and DE-AR0001263 via the Advanced Research Projects Agency - Energy (ARPA-E) program.
\end{acknowledgments}

\section*{Data Availability}

All relevant input files and codes to reproduce the data and results presented in this work can be found on Github \cite{url1,url2}.

\section*{References}
\bibliography{aipsamp}

\end{document}
%